# Some new results concerning the vacuum in Dirac Hole Theory

by

Dan Solomon

Rauland-Borg Corporation
3450 W Oakton
Skokie, IL 60076
USA

Email: dan.solomon@rauland.com

July 3, 2006

**Abstract**

In Dirac's hole theory the vacuum state is generally believed to be the state of minimum energy. It will be shown that this not, in fact, the case and that there must exist states in hole theory with less energy than the vacuum state. It will be shown that energy can be extracted from the hole theory vacuum state through application of an electric field.

# 1. Introduction.

It is well know that there are both positive and negative energy solutions to the Dirac equation. This creates a problem in that an electron in a positive energy state will quickly decay into a negative energy state in the presence of perturbations. This, of course, is not normally observed to occur. Dirac resolved this difficulty by assuming that all the negative energy states are occupied by a single electron and then evoking the Pauli exclusion principle to prevent the decay of a positive energy electron into negative energy states. The electrons in the negative energy states, the so called *Dirac sea*, are assumed to be unobservable. What we observe are variations from the unperturbed vacuum state.

The proposition that the negative energy states are all occupied turns a one electron theory into an N-electron theory where $N \to \infty$. Due to the fact that the negative energy vacuum electrons obey the same Dirac equation as the positive energy electrons we have to, in principle, track the time evolution of an infinite number of states. Also the vacuum electrons in their unperturbed state are unobservable. All we can observe is differences from the unperturbed vacuum state. To simplify the problem quantum field theory was developed. In this case the vacuum state is defined as the state in which no electrons or positrons exist. Therefore at the conceptual level the situation is simplified.

Now consider a "simple" quantum theory consisting of non-interacting electrons in a background classical electromagnetic field. In this case it is generally assumed that quantum field theory (QFT) and hole theory (HT) will give identical results. However, recently, several papers have appeared in the literature pointing out that there are differences between HT and QFT for this simple situation (see [1][2][3][4]). The problem was originally examined by Coutinho et al[1][2]. They calculated the second order change in the energy of the vacuum state due to a time independent perturbation. They found that HT and QFT produce different results. They concluded that the difference between HT and QFT was related to the validity of Feynman's belief that the Pauli Exclusion Principle can be disregarded for intermediate states in perturbation theory. This belief was based on Feynman's observation that terms that violate the Pauli principle formally cancel out in perturbation theory. However Coutino et al show that this is not necessarily the case for HT when applied to an actual problem. This author (Solomon [4]) found that this problem was related to the way the vacuum state was defined in QFT. If the definition of the vacuum state was modified as described in [4] then the HT and QFT would yield identical results.

The above mentioned papers looked at the differences between QFT and HT for a time independent perturbation. A recent paper by this author (Solomon [5]) examined the difference for a particular time dependent perturbation. The conclusion of [5] was that the HT vacuum state is not a minimum energy state, that is, in HT there exist states with less energy than the vacuum state. This is a somewhat surprising result because it is generally assumed that the HT vacuum state is also a minimum energy state. This is in sharp contrast to the standard formulation of QFT where the vacuum state is also the minimum energy state.

One possible result of this research is that it could shed light on the reason for anomalies in QFT. An anomaly occurs when the result of a calculation is not consistent with a symmetry of the Dirac equation. In [5] it was pointed out that HT is anomaly free.

Therefore an understanding of the differences between HT and QFT could help in understanding why anomalies occur in QFT.

In this paper we will examine the fact, which was originally derived in [5], that the HT vacuum state in not the minimum energy state. However the approach used here is significantly different from that of [5]. Therefore we have two different methods of demonstrating that the HT vacuum is not a state of minimum energy. This is important because it corrects the widespread misperception that the HT vacuum is the state of minimum energy.

In the following discussion we will apply a specified electric field to a system that is initially in the HT vacuum state. The vacuum electrons are perturbed by the electric field according to the Dirac equation. The change in energy of each vacuum electron can be calculated. For the specific electric field used in this paper the change in energy of each vacuum electron will be shown to be negative. This means that the total change in the vacuum energy will be negative since the change in the vacuum energy is the sum of the change in the energy of each vacuum electron. Therefore energy has been extracted from the vacuum state producing a new state whose energy is negative with respect to the initial unperturbed vacuum state.

## 2. The Dirac Equation.

In order to simplify the discussion and avoid unnecessary mathematical details we will work in 1-1 dimensional space-time where the space dimension is taken along the z-axis and use natural units so that $\hbar = c = 1$. In this case the Dirac equation for a single electron in the presence of an external electric potential is,

$$i\frac{\partial \psi(z,t)}{\partial t} = H\psi(z,t) \tag{2.1}$$

where the Dirac Hamiltonian is given by,

$$H = H_0 + qV \tag{2.2}$$

where $H_0$ is the Hamiltonian in the absence of interactions, V is an external potential, and q is the electric charge. For the 1-1D case,

$$\hat{H}_0 = \left(-i\sigma_x \frac{\partial}{\partial z} + m\sigma_z\right) \tag{2.3}$$

where $\sigma_x$ and $\sigma_z$ are the usual Pauli matrices. We will assume periodic boundary conditions so that the solutions satisfy $\psi(z,t) = \psi(z+L,t)$ where L is the 1-dimensional integration volume. In this case the orthonormal free field solutions (V is zero) of (2.1) are given by,

$$\varphi_{\lambda,r}^{(0)}(z,t) = \varphi_{\lambda,r}^{(0)}(z)e^{-i\varepsilon_{\lambda,r}^{(0)}t} = u_{\lambda,r}e^{-i\left(\varepsilon_{\lambda,r}^{(0)}t - p_r z\right)} \tag{2.4}$$

where 'r' is an integer, $\lambda = \pm 1$ is the sign of the energy, $p_r = 2\pi r/L$, and where,

$$\varepsilon_{\lambda,r}^{(0)} = \lambda E_r; \quad E_r = +\sqrt{p_r^2 + m^2}; \quad u_{\lambda,r} = N_{\lambda,r}\begin{pmatrix} 1 \\ p_r/(\lambda E_r + m) \end{pmatrix}; \quad N_{\lambda,r} = \sqrt{\frac{\lambda E_r + m}{2L\lambda E_r}} \tag{2.5}$$

The quantities $\varphi_{\lambda,r}^{(0)}(z)$ satisfy the relationship,

$$\hat{H}_0 \varphi_{\lambda,r}^{(0)}(z) = \varepsilon_{\lambda,r}^{(0)} \varphi_{\lambda,r}^{(0)}(z) \tag{2.6}$$

The $\varphi_{\lambda,r}^{(0)}(z)$ form an orthonormal basis set and satisfy,

$$\int_{-L/2}^{+L/2} \varphi_{\lambda,r}^{(0)\dagger}(z)\varphi_{\lambda',s}^{(0)}(z)\,dz = \delta_{\lambda\lambda'}\delta_{rs} \tag{2.7}$$

The energy $\xi(\psi(z,t))$ of a normalized wave function $\psi(z,t)$ is given by,

$$\xi(\psi(z,t)) = \int_{-L/2}^{+L/2} \psi^\dagger(z,t)(H_0 + qV)\psi(z,t)dz \qquad (2.8)$$

## 2. Hole Theory

The proposition that the negative energy states are all occupied turns a one electron theory into an N-electron theory where $N \to \infty$. In this paper we will assume that these electrons are non-interacting.

For an N-electron theory the wave function is written as a Slater determinant [6,7,8],

$$\Psi^N(z_1,z_2,...,z_N,t) = \frac{1}{\sqrt{N!}} \sum_P (-1)^s P(\psi_1(z_1,t)\psi_2(z_2,t)\cdots\psi_N(z_N,t)) \qquad (3.1)$$

where the $\psi_n(z,t)$ ($n=1,2,\ldots,N$) are a normalized and orthogonal set of wave functions that obey the Dirac equation, P is a permutation operator acting on the space coordinates, and s is the number of interchanges in P. Note if $\psi_a(z,t)$ and $\psi_b(z,t)$ are two wave functions that obey the Dirac equation then it can be shown that,

$$\frac{\partial}{\partial t} \int_{-L/2}^{+L/2} \psi_a^\dagger(z,t)\psi_b(z,t)dz = 0 \qquad (3.2)$$

Therefore if the $\psi_n(z,t)$ in (3.1) are orthogonal at some initial time then they are orthogonal for all time.

The expectation value of a single particle operator $O_{op}(z)$ is defined as,

$$O_e = \int \psi^\dagger(z,t) O_{op}(z)\psi(z,t)dz \qquad (3.3)$$

where $\psi(z,t)$ is a normalized single particle wave function. The N-electron operator is given by,

$$O_{op}^N(z_1, z_2, ..., z_N) = \sum_{n=1}^{N} O_{op}(z_n) \qquad (3.4)$$

which is just the sum of one particle operators. The expectation value of a normalized N-electron wave function is,

$$O_e^N = \int \Psi^{N\dagger}(z_1, z_2, ..., x_N, t) O_{op}^N(z_1, z_2, ..., z_N) \Psi^N(z_1, z_2, ..., z_N, t) dz_1 dz_2 ... dz_N \qquad (3.5)$$

This can be shown to be equal to,

$$O_e^N = \sum_{n=1}^{N} \int \psi_n^\dagger(z,t) O_{op}(z) \psi_n(z,t) d\vec{x} \qquad (3.6)$$

That is, the N electron expectation value is just the sum of the single particle expectation values associated with each of the individual wave functions $\psi_n$. For example, the energy $\xi(\Psi^N)$ of the N-electron state is,

$$\xi(\Psi^N) = \sum_{n=1}^{N} \int \psi_n^\dagger(z,t)(H_0 + qV) \psi_n(z,t) dz = \sum_{n=1}^{N} \xi(\psi_n) \qquad (3.7)$$

### 4. Time varying perturbation

In this section we will examine the effect of a time varying electric potential on the HT vacuum state. Assume, at some initial time $t_0$, that the electric potential is zero and that the system is in the unperturbed vacuum state. This is the state where each negative energy wave function $\varphi_{-1,r}^{(0)}$ is occupied by a single electron and each positive energy state $\varphi_{+1,r}^{(0)}$ is unoccupied. Next, consider the change in the energy due to an interaction with an external electric potential which is applied at some time $t > t_0$ and then removed at some later time $t_1$ so that,

$$V(z,t) = 0 \text{ for } t < t_0; \quad V(z,t) \neq 0 \text{ for } t_0 \leq t \leq t_1; \quad V(z,t) = 0 \text{ for } t > t_1 \quad (4.1)$$

Now what is the change in the energy of the quantum system due to this interaction with the electric potential? Under the action of the electric potential each initial wave function $\varphi_{\lambda,r}^{(0)}(z,t_0)$ evolves according to the Dirac equation into the final state $\varphi_{\lambda,r}(z,t_f)$ where $t_f > t_1$. The change in the energy of the state $\varphi_{\lambda,r}^{(0)}(z,t_0)$ from $t_0$ to $t_f$ is given by,

$$\delta\varepsilon_{\lambda,r} = \left\langle \varphi_{\lambda,r}^\dagger(z,t_f) H_0 \varphi_{\lambda,r}(z,t_f) \right\rangle - \varepsilon_{\lambda,r}^{(0)} \quad (4.2)$$

where, to simplify notation, we define $\left\langle g(z) \right\rangle \equiv \int_{-L/2}^{+L/2} g(z) dz$ and we have used the fact that, according to (4.1), $V(z,t_f) = 0$ since $t_f > t_1$. The change in the energy of the vacuum state is the sum of change in energy of each vacuum electron and is therefore given by,

$$\Delta\xi_{vac} = \sum_r \delta\varepsilon_{-1,r} \quad (4.3)$$

In order to calculate $\Delta\xi_{vac}$ it is necessary to calculate $\delta\varepsilon_{-1,r}$ for all r. Since it is not possible to find exact solutions to the Dirac equation for most time dependent potentials we will work this problem using perturbation theory.

The relationship between the initial and final wave function is,

$$\varphi_{\lambda,r}(z,t_f) = U(t_f,t_0) \varphi_{\lambda,r}^{(0)}(z,t_0) \quad (4.4)$$

where $U(t_f, t_0)$ is a unitary operator. From Thaller [9] a formal expression for $U(t_f, t_0)$ is,

$$U(t_f, t_0) = e^{-iH_0 t_f} \left( 1 - iq \int_{t_0}^{t_f} V_I(z,t) dt - q^2 \int_{t_0}^{t_f} V_I(z,t) dt \int_{t_0}^{t} V_I(z,t') dt' + O(q^3) \right) e^{+iH_0 t_0}$$

(4.5)

where $V_I(z,t) = e^{+iH_0 t} V(z,t) e^{-iH_0 t}$ and where $O(q^3)$ means terms to the third order of q or higher. From this we can write,

$$\varphi_{\lambda,r}(z, t_f) = \varphi_{\lambda,r}^{(0)}(z, t_f) + q \varphi_{\lambda,r}^{(1)}(z, t_f) + q^2 \varphi_{\lambda,r}^{(2)}(z, t_f) + O(q^3) \qquad (4.6)$$

The terms $\varphi_{\lambda,r}^{(n)}$ in this expression are determined by using (4.5) in (4.4) and equating the terms that are to the same order in q. Similarly $\delta\varepsilon_{\lambda,r}$ can be expanded as,

$$\delta\varepsilon_{\lambda,r} = q \delta\varepsilon_{\lambda,r}^{(1)} + q^2 \delta\varepsilon_{\lambda,r}^{(2)} + O(q^3) \qquad (4.7)$$

It is shown in Appendix A that the above relationships yield,

$$\delta\varepsilon_{\lambda,r}^{(1)} = 0 \qquad (4.8)$$

and,

$$\delta\varepsilon_{\lambda,r}^{(2)} = \sum_{\lambda'=\pm 1} \sum_{s=-\infty}^{+\infty} \left( |f_{\lambda',s;\lambda,r}|^2 \left( \varepsilon_{\lambda',s}^{(0)} - \varepsilon_{\lambda,r}^{(0)} \right) \right) \qquad (4.9)$$

where,

$$f_{\lambda',s;\lambda,r} = \int_{t_0}^{t_f} V_{\lambda',s;\lambda,r}(t) e^{i(\varepsilon_{\lambda',s} - \varepsilon_{\lambda,r})t} dt \qquad (4.10)$$

and,

$$V_{\lambda',s;\lambda,r}(t) = \left\langle \varphi_{\lambda',s}^{(0)\dagger}(z) V(z,t) \varphi_{\lambda,r}^{(0)}(z) \right\rangle \tag{4.11}$$

## 5. Calculating the change in energy.

We shall now apply the results of the last section to a specific perturbation. Let the electric potential $V(z,t)$ be given by,

$$V(z,t) = 4\cos(k_w z)\left(\frac{\sin(mt)}{t}\right) = \left(e^{ik_w z} + e^{-ik_w z}\right)\int_{-m}^{+m} e^{iqt} dq \tag{5.1}$$

where m is the mass of the electron and $k_w = 2\pi w/L < m$ where w is a positive integer. It is obvious from the above expression that $V(z,t) \to 0$ at $t \to \pm\infty$. Under the action of this electric potential each initial wave function $\varphi_{\lambda,r}^{(0)}(z,t_0)$, where $t_0 \to -\infty$, evolves into the final wave function $\varphi_{\lambda,r}(z,t_f)$ where $t_f \to +\infty$. Use (5.1) in (4.10) to obtain,

$$f_{\lambda',s;\lambda,r} = \left\langle \varphi_{\lambda',s}^{(0)\dagger}(z)\left(e^{ik_w z} + e^{-ik_w z}\right)\varphi_{\lambda,r}^{(0)}(z) \right\rangle \int_{-\infty}^{+\infty} e^{i(\varepsilon_{\lambda',s} - \varepsilon_{\lambda,r})t}\left(\int_{-m}^{+m} e^{iqt} dq\right) dt \tag{5.2}$$

Use (2.5) in the above to obtain,

$$f_{\lambda',s;\lambda,r} = u_{\lambda',s}^{\dagger} u_{\lambda,r} \int_{-\infty}^{+\infty} dt \int_{-L/2}^{+L/2} dz \begin{pmatrix} e^{-i(\lambda E_r - \lambda' E_s)t} e^{i(p_r - p_s)z} \\ \times \left(e^{ik_w z} + e^{-ik_w z}\right)\int_{-m}^{+m} e^{iqt} dq \end{pmatrix} \tag{5.3}$$

Perform the integrations over t and z to obtain,

$$f_{\lambda',s;\lambda,r} = 2\pi \left\{ u_{\lambda',s}^{\dagger} u_{\lambda,r} \begin{pmatrix} \delta_L(w+r-s) \\ +\delta_L(-w+r-s) \end{pmatrix} \int_{-m}^{+m} \delta\begin{pmatrix} -\lambda E_r \\ +\lambda' E_s + q \end{pmatrix} dq \right\} \tag{5.4}$$

where,

$$\delta_L(r) = \begin{cases} L & \text{if } r = 0 \\ 0 & \text{if } r \neq 0 \end{cases} \tag{5.5}$$

12From the definition of the delta function we have the following relationship

$$\delta_L\left(\pm w+r-s\right)\int_{-m}^{+m}\delta\begin{pmatrix}-\lambda E_r\\+\lambda'E_s+q\end{pmatrix}dq=\delta_L\left(\pm w+r-s\right)\int_{-m}^{+m}\delta\begin{pmatrix}-\lambda E_{s\mp w}\\+\lambda'E_s+q\end{pmatrix}dq \quad (5.6)$$

Due to the fact that $(E_r+E_s)\geq 2m$ and $k_w<m$ we have that,

$$\int_{-m}^{+m}\delta\left(-\lambda E_{s\mp w}+\lambda'E_s+q\right)dq=\delta_{\lambda\lambda'} \quad (5.7)$$

Use these results in (5.4) to obtain,

$$f_{\lambda',s;\lambda,r}=2\pi\begin{pmatrix}u^\dagger_{\lambda',r+w}u_{\lambda,r}\delta_L(w+r-s)\\+u^\dagger_{\lambda',r-w}u_{\lambda,r}\delta_L(-w+r-s)\end{pmatrix}\delta_{\lambda\lambda'} \quad (5.8)$$

This yields,

$$\left|f_{\lambda',s;\lambda,r}\right|^2=4\pi^2 L\begin{pmatrix}\left|u^\dagger_{\lambda',r+w}u_{\lambda,r}\right|^2\delta_L(w+r-s)\\+\left|u^\dagger_{\lambda',r-w}u_{\lambda,r}\right|^2\delta_L(-w+r-s)\end{pmatrix}\delta_{\lambda\lambda'} \quad (5.9)$$

where we have used $(\delta_{\lambda\lambda'})^2=\delta_{\lambda\lambda'}$, $(\delta_L(r))^2=L\delta_L(r)$, and

$\delta_L(w+r-s)\delta_L(-w+r-s)=0$. Use (5.9) in (4.9) to obtain,

$$\delta\varepsilon^{(2)}_{\lambda,r}=4\pi^2L^2\begin{pmatrix}\left|u^\dagger_{\lambda,r+w}u_{\lambda,r}\right|^2\left(\varepsilon^{(0)}_{\lambda,r+w}-\varepsilon^{(0)}_{\lambda,r}\right)\\+\left|u^\dagger_{\lambda,r-w}u_{\lambda,r}\right|^2\left(\varepsilon^{(0)}_{\lambda,r-w}-\varepsilon^{(0)}_{\lambda,r}\right)\end{pmatrix} \quad (5.10)$$

It is shown in Appendix B that,

$$\delta\varepsilon^{(2)}_{\lambda,r}=2\pi^2\lambda k_w\left(\frac{(p_r+k_w)}{E_{r+w}}-\frac{(p_r-k_w)}{E_{r-w}}\right) \quad (5.11)$$

Therefore, for negative energy states $(\lambda=-1)$,

$$\delta\varepsilon_{-1,r}^{(2)} = -2\pi^2 k_w \left( \frac{(p_r + k_w)}{E_{r+w}} - \frac{(p_r - k_w)}{E_{r-w}} \right) < 0 \tag{5.12}$$

where it is shown in Appendix C that this quantity is negative for all $\delta\varepsilon_{-1,r}^{(2)}$. Using this result along with (4.8) in (4.7), we obtain,

$$\delta\varepsilon_{-1,r} = q^2 \delta\varepsilon_{-1,r}^{(2)} + O(q^3) \underset{q \to 0}{=} q^2 \delta\varepsilon_{-1,r}^{(2)} < 0 \tag{5.13}$$

Therefore, for sufficiently small q, the change in the energy of each vacuum electron is negative. (Note, in Dirac equation the quantity q always appears multiplying the electric potential, i.e., as qV. Therefore the limit $q \to 0$ is equivalent to holding q fixed and letting $V \to 0$.) The change in the vacuum energy is simply the sum of the change in the energy of each vacuum electron. Therefore the change in the vacuum energy must be negative. This means that energy was extracted from the vacuum state due to application of the electric field.

To calculate the total change to the second order in the vacuum energy, $\Delta\xi_{vac}^{(2)}$, we must sum the quantities $q^2 \delta\varepsilon_{-1,r}^{(2)}$ over all values of the index r. If we let the integration length $L \to \infty$ we can integrate over the momentum $p = 2\pi r/L$ to obtain,

$$\Delta\xi_{vac}^{(2)} = -\pi q^2 kL \int_{-\infty}^{+\infty} \left( \frac{(p+k)}{E_{p+k}} - \frac{(p-k)}{E_{p-k}} \right) dp \tag{5.14}$$

where, in the above expression, $E_{(p \pm k)} = \sqrt{(p \pm k)^2 + m^2}$ and we write k instead of $k_w$.

Now before evaluating this integral let us examine the integrand at large p. At large p it can be shown the integrad drops off faster than $(1/p^2)$. Therefore the integral will be convergent. To evaluate this rewrite (5.14) as,

$$\Delta \xi_{vac}^{(2)} \underset{B \to \infty}{=} -\pi q^2 kL \int_{-B}^{+B} \left( \frac{(p+k)}{E_{p+k}} - \frac{(p-k)}{E_{p-k}} \right) dp \qquad (5.15)$$

This is readily evaluated as,

$$\Delta \xi_{vac}^{(2)} \underset{B \to \infty}{=} -\pi q^2 kL \left( \left( E_{B+k} - E_{-B+k} \right) - \left( E_{B-k} - E_{-B-k} \right) \right) \qquad (5.16)$$

This yields,

$$\Delta E_{vac}^{(2)} \underset{B \to \infty}{=} -2\pi q^2 kL \left( E_{B+k} - E_{B-k} \right) \qquad (5.17)$$

For large B we obtain,

$$E_{(B \pm k)} \underset{B \to \infty}{=} (B \pm k) \left( 1 + \frac{m^2}{2(B \pm k)^2} + \cdots \right) \qquad (5.18)$$

Use this in (5.17) to obtain,

$$\Delta \xi_{vac}^{(2)} = -4\pi q^2 k^2 L \qquad (5.19)$$

Therefore the second order change in the vacuum energy is a well defined negative quantity. The total change in the vacuum energy is obviously,

$$\Delta \xi_{vac} = -4\pi q^2 k^2 L + O(q^3) \underset{q \to 0}{=} -4\pi q^2 k^2 L$$

## 6. Conclusion

We have examined the vacuum in Dirac's hole theory. The vacuum electrons obey the Dirac equation and the energy of these electrons will change in response to an applied electric field. It has been shown that it is possible to find an electric field for which the change in the energy of each vacuum electron is negative. Therefore the total change in the energy of the vacuum state is negative. This new state will have less energy than the original unperturbed vacuum state.

### **Appendix A**

Since the Dirac equation does not affect the normalization condition we have,

$$\left\langle \varphi^{\dagger}_{\lambda,r}(z,t_f)\varphi_{\lambda,r}(z,t_f)\right\rangle = \left\langle \varphi^{(0)\dagger}_{\lambda,r}(z,t_f)\varphi^{(0)}_{\lambda,r}(z,t_f)\right\rangle = 1 \qquad (A.1)$$

Use (4.6) in the above to obtain,

$$0 = q\left(\left\langle \varphi^{(0)\dagger}_{\lambda,r}\varphi^{(1)}_{\lambda,r}\right\rangle + \left\langle \varphi^{(1)\dagger}_{\lambda,r}\varphi^{(0)}_{\lambda,r}\right\rangle\right) + q^2\left(\left\langle \varphi^{(0)\dagger}_{\lambda,r}\varphi^{(2)}_{\lambda,r}\right\rangle + \left\langle \varphi^{(2)\dagger}_{\lambda,r}\varphi^{(0)}_{\lambda,r}\right\rangle + \left\langle \varphi^{(1)\dagger}_{\lambda,r}\varphi^{(1)}_{\lambda,r}\right\rangle\right) + O(q^3)$$

(A.2)

This yields,

$$q^2\left\langle \varphi^{(1)\dagger}_{\lambda,r}\varphi^{(1)}_{\lambda,r}\right\rangle = -q\left(\left\langle \varphi^{(0)\dagger}_{\lambda,r}\varphi^{(1)}_{\lambda,r}\right\rangle + \left\langle \varphi^{(1)\dagger}_{\lambda,r}\varphi^{(0)}_{\lambda,r}\right\rangle\right) - q^2\left(\left\langle \varphi^{(0)\dagger}_{\lambda,r}\varphi^{(2)}_{\lambda,r}\right\rangle + \left\langle \varphi^{(2)\dagger}_{\lambda,r}\varphi^{(0)}_{\lambda,r}\right\rangle\right) + O(q^3)$$

(A.3)

Next use (4.2) and (4.6) in (4.7) to obtain,

$$\delta\varepsilon_{\lambda,r} = \left\langle \varphi^{(0)\dagger}_{\lambda,r}H_0\varphi^{(0)}_{\lambda,r}\right\rangle + q\left(\begin{array}{c}\left\langle \varphi^{(1)\dagger}_{\lambda,r}H_0\varphi^{(0)}_{\lambda,r}\right\rangle \\ +\left\langle \varphi^{(0)\dagger}_{\lambda,r}H_0\varphi^{(1)}_{\lambda,r}\right\rangle\end{array}\right) + q^2\left(\begin{array}{c}\left\langle \varphi^{(2)\dagger}_{\lambda,r}H_0\varphi^{(0)}_{\lambda,r}\right\rangle + \left\langle \varphi^{(1)\dagger}_{\lambda,r}H_0\varphi^{(1)}_{\lambda,r}\right\rangle \\ +\left\langle \varphi^{(0)\dagger}_{\lambda,r}H_0\varphi^{(2)}_{\lambda,r}\right\rangle\end{array}\right) + O(q^3) - \varepsilon^{(0)}_{\lambda,r}$$

(A.4)

Use (2.6) in the above to obtain,

$$\delta\varepsilon_{\lambda,r} = q\varepsilon^{(0)}_{\lambda,r}\left(\begin{array}{c}\left\langle \varphi^{(1)\dagger}_{\lambda,r}\varphi^{(0)}_{\lambda,r}\right\rangle \\ +\left\langle \varphi^{(0)\dagger}_{\lambda,r}\varphi^{(1)}_{\lambda,r}\right\rangle\end{array}\right) + q^2\varepsilon^{(0)}_{\lambda,r}\left(\begin{array}{c}\left\langle \varphi^{(2)\dagger}_{\lambda,r}\varphi^{(0)}_{\lambda,r}\right\rangle \\ +\left\langle \varphi^{(0)\dagger}_{\lambda,r}\varphi^{(2)}_{\lambda,r}\right\rangle\end{array}\right) + q^2\left\langle \varphi^{(1)\dagger}_{\lambda,r}H_0\varphi^{(1)}_{\lambda,r}\right\rangle + O(q^3) \quad (A.5)$$

Use (A.3) in the above to obtain,

$$\delta\varepsilon_{\lambda,r} = q^2\left\langle \varphi^{(1)\dagger}_{\lambda,r}H_0\varphi^{(1)}_{\lambda,r}\right\rangle - q^2\varepsilon^{(0)}_{\lambda,r}\left\langle \varphi^{(1)\dagger}_{\lambda,r}\varphi^{(1)}_{\lambda,r}\right\rangle + O(q^3) \qquad (A.6)$$

Therefore,

$$\delta\varepsilon^{(1)}_{\lambda,r} = 0 \tag{A.7}$$

and

$$\delta\varepsilon^{(2)}_{\lambda,r} = \left\langle \varphi^{(1)\dagger}_{\lambda,r}(z,t_f) H_0 \varphi^{(1)}_{\lambda,r}(z,t_f) \right\rangle - \varepsilon^{(0)}_{\lambda,r} \left\langle \varphi^{(1)\dagger}_{\lambda,r}(z,t_f) \varphi^{(1)}_{\lambda,r}(z,t_f) \right\rangle \tag{A.8}$$

Evaluate this equation as follows. From (4.4), (4.5), and (4.6) we obtain,

$$\varphi^{(1)}_{\lambda,r}(z,t_f) = -i \int_{t_0}^{t_f} e^{-iH_0(t_f-t)} V(z,t) e^{-iH_0(t-t_0)} \varphi^{(0)}_{\lambda,r}(z,t_0) dt \tag{A.9}$$

This becomes,

$$\varphi^{(1)}_{\lambda,r}(z,t_f) = -i \int_{t_0}^{t_f} e^{-iH_0(t_f-t)} V(z,t) \varphi^{(0)}_{\lambda,r}(z,t) dt \tag{A.10}$$

Define,

$$V_{\lambda',s;\lambda,r}(t) = \left\langle \varphi^{(0)\dagger}_{\lambda',s}(z) V(z,t) \varphi^{(0)}_{\lambda,r}(z) \right\rangle \tag{A.11}$$

Since the $\varphi^{(0)}_{\lambda,r}(z)$ form on orthonormal basis we can expand the quantity $V(z,t)\varphi^{(0)}_{\lambda,r}(z)$ as a Fourier expansion as follows,

$$V(z,t)\varphi^{(0)}_{\lambda,r}(z) = \sum_{\lambda'=\pm 1} \sum_{s=-\infty}^{+\infty} \varphi^{(0)}_{\lambda',s}(z) V_{\lambda',s;\lambda,r}(t) \tag{A.12}$$

Use this in (A.10) to obtain,

$$\varphi^{(1)}_{\lambda,r}(z,t_f) = -i \sum_{\lambda'=\pm 1} \sum_{s=-\infty}^{+\infty} \varphi^{(0)}_{\lambda',s}(z,t_f) f_{\lambda',s;\lambda,r} \tag{A.13}$$

where,

$$f_{\lambda',s;\lambda,r} = \int_{t_0}^{t_f} V_{\lambda',s;\lambda,r}(t) e^{i\left(\varepsilon^{(0)}_{\lambda',s}-\varepsilon^{(0)}_{\lambda,r}\right)t} dt \tag{A.14}$$

Use this in (A.8) to obtain,

$$\delta\varepsilon_{\lambda,r}^{(2)} = \sum_{\lambda'=\pm 1}\sum_{s=-\infty}^{+\infty}\left(\left|f_{\lambda',s;\lambda,r}\right|^2\left(\varepsilon_{\lambda',s}^{(0)}-\varepsilon_{\lambda,r}^{(0)}\right)\right) \qquad (A.15)$$

## **Appendix B**

We want to evaluate equation (5.10) and show that it results in equation (5.11). Use (2.5) to obtain,

$$u_{\lambda,r\pm w}^\dagger u_{\lambda,r} = \sqrt{\frac{\lambda E_{r\pm w}+m}{2\lambda L E_{r\pm w}}}\sqrt{\frac{\lambda E_r+m}{2\lambda L E_r}}\left(1+\frac{p_r(p_r+k)}{(\lambda E_{r\pm w}+m)(\lambda E_r+m)}\right) \qquad (B.1)$$

Use $(\lambda E_r)^2 - m^2 = p_r^2$ in the above to obtain,

$$u_{\lambda,r\pm w}^\dagger u_{\lambda,r} = \sqrt{\frac{\lambda E_{r\pm w}+m}{2\lambda L E_{r\pm w}}}\sqrt{\frac{\lambda E_r+m}{2\lambda L E_r}}\left(1+\frac{(\lambda E_{r\pm w}-m)(\lambda E_r-m)}{p_r(p_r\pm k_w)}\right) \qquad (B.2)$$

Use this result to yield,

$$\left|u_{\lambda,r\pm w}^\dagger u_{\lambda,r}\right|^2 = \left(\frac{(\lambda E_{r\pm w})(\lambda E_r)+p_r(p_r\pm k_w)+m^2}{2(\lambda E_{r\pm w})(\lambda E_r)L^2}\right) \qquad (B.3)$$

Use this in (5.10) along with the fact that $\lambda^2 = 1$ to obtain

$$\delta\varepsilon_{\lambda,r}^{(2)} = 2\pi^2\lambda\left(\begin{aligned}&\left(1+\frac{p_r(p_r+k_w)+m^2}{E_{r+w}E_r}\right)(E_{r+w}-E_r)\\&+\left(1+\frac{p_r(p_r-k_w)+m^2}{E_{r-w}E_r}\right)(E_{r-w}-E_r)\end{aligned}\right) \qquad (B.4)$$

This yields,

$$\delta\varepsilon_{\lambda,r}^{(2)} = 2\pi^2\lambda\left(\begin{aligned}&\left(1+\frac{E_{r+w}}{E_r}-\frac{k(p_r+k_w)}{E_{r+w}E_r}\right)(E_{r+w}-E_r)\\&+(w\to -w)\end{aligned}\right) \qquad (B.5)$$

Some additional algebraic manipulation yields,

$$\delta\varepsilon_{\lambda,r}^{(2)} = 2\pi^2\lambda \left( \left( -E_r + \frac{E_{r+w}^2}{E_r} - \frac{k_w(p_r + k_w)}{E_r} + \frac{k_w(p_r + k_w)}{E_{r+w}} \right) + (w \to -w) \right) \quad (B.6)$$

Use some simple algebra to obtain,

$$\delta\varepsilon_{\lambda,r}^{(2)} = 2\pi^2\lambda \left( \left( \frac{p_r k_w}{E_r} + \frac{k_w(p_r + k_w)}{E_{r+w}} \right) + \left( \frac{-p_r k_w}{E_r} - \frac{k_w(p_r - k_w)}{E_{r-w}} \right) \right) \quad (B.7)$$

Use this result to yield (5.11).

## **Appendix C**

Assume $k_w$ is positive. Then it can be shown that,

$$\frac{(p_r + k_w)}{E_{r+w}} > \frac{(p_r - k_w)}{E_{r-w}} \quad \text{for all } p_r \quad (C.1)$$

First consider the case where $p_r$ is positive. The relationship is obviously true for $k_w > p_r$. Now let $p_r > k_w$. In this case both sides of (C.1) are positive therefore we can square both sides to obtain,

$$(p_r + k_w)^2 E_{r-w}^2 > (p_r - k_w)^2 E_{r+w}^2 \quad (C.2)$$

From this we obtain,

$$(p_r + k_w)^2 \left( (p_r - k_w)^2 + m^2 \right) > (p_r - k_w)^2 \left( (p_r + k_w)^2 + m^2 \right) \quad (C.3)$$

This yields,

$$(p_r + k_w)^2 > (p_r - k_w)^2 \quad (C.4)$$

which is true for positive $k_w$ and $p_r > k_w$. If $p_r$ is negative then (C.1) becomes,

$$\frac{(-|p_r| + k_w)}{E_{|r|-w}} > \frac{(-|p_r| - k_w)}{E_{|r|+w}} \quad (C.5)$$

This yields,

$$\frac{\left(|p_r|+k_w\right)}{E_{|r|+w}} > \frac{\left(|p_r|-k_w\right)}{E_{|r|-w}} \qquad (C.6)$$

which is obviously true from the previous discussion.

## **References**

1. F.A.B. Coutinho, D. Kaing, Y. Nagami, and L. Tomio, Can. J. of Phys., **80**, 837 (2002). (see also quant-ph/0010039).

2. F.A.B. Coutinho, Y. Nagami, and L. Tomio, Phy. Rev. A, **59**, 2624 (1999).

3. R. M. Cavalcanti, quant-ph/9908088.

4. D. Solomon. Can. J. Phys., **81**, 1165, (2003).

5. D. Solomon. Can. J. Phys., **83**, 257, (2005)

6. E.K.U. Gross, E. Runge, O. Heinonen, "Many Particle Theory", Adam Hilger, Bristol (1991).

7. P. Roman, "Advanced Quantum Theory", Addison-Wesly Publishing Co., Inc., Reading, Massachusetts, (1965).

8. S. Raines, "Many-Electron Theory", North Holland Publishing Co., Amsterdam (1972).

9. B. Thaller, "The Dirac Equation", Springer-Verlag, Berlin (1992).